\documentclass[twocolumn]{aastex631}
\usepackage{amsmath}

\newcommand{\ha}{\hbox{H$\alpha$}}
\newcommand{\hb}{\hbox{H$\beta$}}
\newcommand{\hd}{\hbox{H$\delta$}}

\newcommand{\oii}{\hbox{[O\,{\sc ii}]}}
\newcommand{\Msun}{\hbox{$M_{\sun}$}}

\shorttitle{size--mass relation of PSB galaxies}
\shortauthors{Chen et al.}
\graphicspath{{figures/}}

\begin{document}

\title{The Size--Mass Relation of Post-Starburst Galaxies in the Local Universe}

\correspondingauthor{Zesen Lin, Xu Kong}
\email{zesenlin@ustc.edu.cn, xkong@ustc.edu.cn}

\author[0000-0002-5016-6901]{Xinkai Chen}
\affiliation{CAS Key Laboratory for Research in Galaxies and Cosmology, Department of Astronomy, University of Science and Technology of China, Hefei 230026, China}

\affiliation{School of Astronomy and Space Sciences, University of Science and Technology of China, Hefei 230026, China}

\author[0000-0001-8078-3428]{Zesen Lin}
\affiliation{CAS Key Laboratory for Research in Galaxies and Cosmology, Department of Astronomy, University of Science and Technology of China, Hefei 230026, China}
\affiliation{School of Astronomy and Space Sciences, University of Science and Technology of China, Hefei 230026, China}

\author[0000-0002-7660-2273]{Xu Kong}
\affiliation{CAS Key Laboratory for Research in Galaxies and Cosmology, Department of Astronomy, University of Science and Technology of China, Hefei 230026, China}
\affiliation{School of Astronomy and Space Sciences, University of Science and Technology of China, Hefei 230026, China}
\affiliation{Frontiers Science Center for Planetary Exploration and Emerging Technologies, University of Science and Technology of China, Hefei, Anhui, 230026, China}

\author{Zhixiong Liang}
\affiliation{CAS Key Laboratory for Research in Galaxies and Cosmology, Department of Astronomy, University of Science and Technology of China, Hefei 230026, China}
\affiliation{School of Astronomy and Space Sciences, University of Science and Technology of China, Hefei 230026, China}

\author{Guangwen Chen}
\affiliation{CAS Key Laboratory for Research in Galaxies and Cosmology, Department of Astronomy, University of Science and Technology of China, Hefei 230026, China}
\affiliation{School of Astronomy and Space Sciences, University of Science and Technology of China, Hefei 230026, China}

\author{Hong-Xin Zhang}
\affiliation{CAS Key Laboratory for Research in Galaxies and Cosmology, Department of Astronomy, University of Science and Technology of China, Hefei 230026, China}

\affiliation{School of Astronomy and Space Sciences, University of Science and Technology of China, Hefei 230026, China}

\begin{abstract}

We present a study of the size--mass relation for local post-starburst (PSB) galaxies at $z\lesssim0.33$ selected from the Sloan Digital Sky Survey Data Release 8. We find that PSB galaxies with stellar mass ($M_*$) at $10^9~\Msun<M_*<10^{12}~\Msun$ have their galaxy size smaller than or comparable with those of quiescent galaxies (QGs). After controlling redshift and stellar mass, the sizes of PSBs are $\sim 13\%$ smaller on average than those of QGs, such differences become larger and significant towards the low-$M_*$ end, especially at $10^{9.5}~\Msun \lesssim M_*\lesssim 10^{10.5}~\Msun$ where PSBs can be on average $\sim 19\%$ smaller than QGs.
In comparison with predictions of possible PSB evolutionary pathways from cosmological simulations, we suggest that a fast quenching of star formation following a short-lived starburst event (might be induced by major merger) should be the dominated pathway of our PSB sample. Furthermore, by cross-matching with group catalogs, we confirm that local PSBs at $M_*\lesssim10^{10}~\Msun$ are more clustered than more massive ones. PSBs resided in groups are found to be slightly larger in galaxy size and more disk-like compared to field PSBs, which is qualitatively consistent with and thus hints the environment-driven fast quenching pathway for group PSBs. Taken together, our results support multiple evolutionary pathways for local PSB galaxies: while massive PSBs are thought of as products of fast quenching following a major merger-induced starburst, environment-induced fast quenching should play a role in the evolution of less massive PSBs, especially at $M_*\lesssim 10^{10}~\Msun$.
\end{abstract}

\keywords{galaxy evolution (594); galaxy structure (622); galaxy quenching (2040); post-starburst galaxies (2176);}

\section{Introduction} \label{sec:intro}

Galaxies in the local Universe show a clear bimodality in the color--magnitude diagram \citep{baldry2004quantifying}, which suggests that they can be naturally separated into two populations: star-forming galaxies (SFGs) in the ``blue cloud'' and quiescent galaxies (QGs) in the ``red sequence''. The dearth of galaxies in the transitional region, which is known as ``green valley'' (GV) region \citep{martin2007uvoptical,salim2007uv,schiminovich2007uvoptical,wyder2007uvoptical,salim2014green}, indicates that the migration from blue cloud to red sequence is relatively rapid \citep{martin2007uvoptical}. Therefore, investigations of intermediate galaxies resided in this transitional region can give insights into galaxy evolution (e.g., \citealt{schawinski2014green,gu2018morphological,suess2021dissecting,lin2022almaquest}).

Post-starburst (PSB) galaxies, also known as E+A/K+A galaxies, have spectral features of elliptical galaxies and A-type stars \citep{dressler1983spectroscopy,poggianti1999star}. These galaxies are observed to reside in the GV region (e.g., \citealt{wong2012galaxy,greene2021refining}), implying that they might be one kind of transitional galaxies. The strong Balmer absorption lines indicate that those galaxies have experienced starbursts within the last 1 Gyr since the lifetime of an A-type star is about 1 Gyr. But they show no sign of ongoing star formation as first indicated by the non-detection of the $\oii$ emission lines \citep{goto2005266}.

As one of the fundamental scaling relations, galaxy size--stellar mass ($M_*$) relation and its evolution with cosmic time are widely used to probe the assembly history of galaxies and thus study galaxy evolution (e.g., \citealt{williams2010evolving,newman2012can,vanderwel20143dhst,shibuya2015morphologies,kawinwanichakij2021hyper}). When $M_*\lesssim 10^{11}~\Msun$, galaxy sizes of SFGs are found to be larger than those of QGs at a fixed $M_*$ \citep{shen2003size,vanderwel20143dhst,kawinwanichakij2021hyper}, while GV galaxies tend to have galaxy sizes in between \citep{salim2014green,gu2018morphological,suess2021dissecting}.

As for PSB galaxies, high-redshift samples were found to have smaller sizes compared to QGs at a fixed $M_*$ \citep{whitaker2012large,yano2016relation}. Particularly, \citet{almaini2017massive} studied PSB galaxies at high redshift ($z>1$) and found that the sizes of PSB galaxies is smaller than that of QGs, they claimed that massive passive galaxies are formed from proto-spheroids, then rapid quenching creates red nuggets with post-starburst features, and finally the gradual growth in size causes the sizes increase. In this physical picture, galaxies may experience violent events, such as galaxy mergers. The process period can be less than a few hundred Myr, and the intense star formation  quickly exhaust gas and build a dense center core. As new stars are formed in the center, fast process can significantly change galaxies structure, causing a few times smaller size in comparison with their progenitor \citep{hopkins2013star,wellons2015formation,wu2018fast}.

However, this evolutionary pathway might be not true for PSB galaxies at all redshifts. On the one hand, based on ultra-deep optical spectra of PSBs at $z \sim 0.8$, studies combining stellar populations and galaxy structure supported a fast quenching pathway associated with centrally concentrated starbursts and thus dramatic structural changes for PSBs at this epoch \citep{wu2018fast,wu2020colors, deugenio2020inverse}. On the other hand, \citet{maltby2018structure} presented a comparison between lower redshift ($0.5<z<1$) and higher redshift ($z>1$) PSB galaxies, in which the authors found that lower redshift PSBs are less concentrated than their high-redshift analogues, suggesting that the processes that are responsible for PSB evolution at intermediate-redshift might be less violent and there may have different quenching routes at different redshifts.

For local PSB galaxies, one might wonder whether the galaxy size--mass relation observed for intermediate- and high-redshift PSBs (i.e., smaller galaxy sizes compared to QGs at a fixed $M_*$) is still applicable, and thus share the same fast quenching picture used to explain the compact morphology of high-redshift PSBs. Observational studies highlighted the diversity of this population and claimed different $M_*$-dependent evolutionary pathways of PSBs including the fast quenching with significant structural changes and cyclic evolution of SFGs or QGs without substantial morphological transformation \citep{pawlik2018origins}. These pathways were also reproduced by cosmological simulations \citep{pawlik2019diverse}. Recent surveys of PSB galaxies in local galaxy groups/clusters further reported a positive correlation between the incidence of PSB galaxies and environment indicators, which is suggestive of a nonnegligible effect of environment in the evolution of PSB galaxies \citep{poggianti2017gasp,paccagnella2019strong}.

Given that structural changes are closely related with the various evolutionary pathways of PSB galaxies in the local Universe \citep{pawlik2018origins,pawlik2019diverse}, the size--mass relation could be an important probe to clarify the evolution of PSBs and, however, is still lacking in direct study. Therefore, we present a study of the size--mass relation for local PSB galaxies based on data from the Sloan Digital Sky Survey Data Release 8 (SDSS DR8; \citealt{aihara2011eighth}).
This paper is organized as follows. In Section \ref{sec:sample} we describe our sample. The main result is presented and studied in Section \ref{sec:result}. We discuss the implication of our findings in Section \ref{sec:discussion} and summarize in Section \ref{sec:summary}. Throughout this paper, we adopt a $\Lambda$CDM cosmology with $\Omega_{m}=0.3$, $\Omega_{\Lambda}=0.7$ and $H_{0}$=70 km s$^{-1}$ Mpc$^{-1}$.

\section{Sample Selection} \label{sec:sample}
\subsection{Post-starburst Galaxies}
\label{subsec:selection}

PSB galaxies are defined by two criteria, one is significant recent star formation, which is indicated by the presence of a large population of short lifetime ($<1$ Gyr) A type stars and thus the observable strong Balmer absorption, the other one is no ongoing star formation, which is traced by weak emission lines (e.g., $\ha$ and/or $\oii$).
However, the use of nebular emission lines to select PSBs may bias the selection against galaxies hosting narrow-line active galactic nuclei or shocks and also exclude galaxies that are PSBs but not fully quenched \citep{yan2006origin,wild2007bursty,wild2009poststarburst,yesuf2014starburst}. On the other hand, not selecting on emission lines may cause some PSBs to be indistinguishable from star-forming population \citep{wild2007bursty}.

Here, we select PSB galaxies based on the MPA-JHU catalog\footnote{\url{http://www.sdss3.org/dr10/spectro/galaxy_mpajhu.php}} for the SDSS DR8 galaxy spectra, adopting the criteria of $\mathrm{\hd_A>4~\textup{\AA}}$, $\mathrm{EW(\ha) < 3~\textup{\AA}}$, and the median spectral signal-to-noise ratio per pixel of $\mathrm{S/N} > 5$. After careful visual inspection of the selected spectra, we further require spectra to meet $\mathrm{\hd_A/\sigma(\hd_A)>3}$ and $\chi^2_{\mathrm{red,H\alpha}} >0$ to ensure reliable measurements of $\hd_{\mathrm{A}}$ and $\mathrm{EW(\ha)}$, respectively.\footnote{Here, $\sigma(\hd_{\mathrm{A}})$ is the uncertainty of index $\hd_{\mathrm{A}}$, while $\chi^2_{\mathrm{red,H\alpha}}$ is the reduced chi-squared of the emission line fitting of \ha.}  22 sources with non-zero SDSS ``zWarning'' bitmask\footnote{\url{https://www.sdss.org/dr16/algorithms/bitmasks/\#ZWARNING}} (i.e., the best-fit classification or redshift might be not reliable) are also examined via spectra, optical images, and cross-matched information in other online database (e.g., SIMBAD; \citealt{wenger2000simbad}), 17 of them are identified as QSOs or white dwarfs and are removed from the sample. Relaxing these criteria to include a larger region on the EW(\ha)--$\hd_{\mathrm{A}}$ plane (e.g., \citealt{chen2019poststarburst}) or applying a stricter cut on $\hd_{\mathrm{A}}$ (e.g., \citealt{goto2007catalogue}) would not change our main results significantly.

To obtain reliable size measurements, we match the sample with the UPenn SDSS PhotDec Catalog (\citealt{meert2015catalogue,meert2016catalogue}; see Section \ref{subsec:determination_para}), which naturally introduces a $r$ band magnitude limit of $14.0~\mathrm{mag }<r< 17.77~\mathrm{mag}$ \citep{meert2015catalogue}. We also require galaxies to have stellar mass ($M_*$; see Section \ref{subsec:determination_para}) between $10^{9}$ to $10^{12}~\Msun$ within which most of the selected PSBs locate. After these criteria, we obtain 1,479 candidates of PSB galaxies. These PSB galaxies have a redshift ranged from 0.012 to 0.324 and with a median and 1-$\sigma$ range of $0.130_{-0.061}^{+0.069}$.

\subsection{Determination of Physical Properties}
\label{subsec:determination_para}

The UPenn SDSS PhotDec Catalog provides point spread function (PSF)-corrected 2D profile fitting results for $\sim 7\times 10^{5}$ galaxies using a variety of models, including de Vaucouleurs, S\'ersic, de Vaucouleurs+Exponential, and S\'ersic+Exponential models. The fitting pipeline is called \texttt{PyMorph} \citep{vikram2010pymorph}, which is a python based automated software pipeline built on \texttt{SExtractor} \citep{bertin1996sextractor} and the 2D fitting routine \texttt{GALFIT} \citep{peng2002detailed}. To be consistent with previous PSB studies (e.g., \citealt{yano2016relation, almaini2017massive,wu2018fast,wu2020colors}) and enable plausible comparisons in the following analyses, we adopt the effective radius ($R_{\rm e}$) measured along the semi-major axis derived from the single S\'ersic model fitting in $r$-band as our fiducial galaxy size measurement. This choice of $R_{\rm e}$ is generally consistent with most of PSB literature we used for comparison in terms of both the derivation method and the sampled rest-frame wavelengths (see Section \ref{subsec:influ_re} for more details). The statistical uncertainties of size measurements of our PSB sample have a median value and 1-$\sigma$ range of $3.5_{-1.9}^{+4.1}$\%. While the systematic uncertainties, together with the PSF effects, of the size measurement will be discussed in Section \ref{subsec:influ_re}.

The total stellar mass is taken from the MPA-JHU catalog, which is the median value of the probability distribution function (PDF) of the total $M_*$ constructed from the $ugriz$ galaxy photometry and a large library of spectral energy distribution (SED) models sampled the full plausible ranges of galaxy physical properties \citep{kauffmann2003stellar}. To generate model SEDs, a \cite{kroupa2001variation} initial mass function, a \cite{CF2000dust} dust attenuation curve, and the \cite{BC03MNRAS} population synthesis code are adopted, while star formation history is assumed to be an exponentially declining continuous star formation superimposed with random bursts. The PDF of each galaxy parameter (e.g., $M_*$) is computed via weighting each model by the corresponding $\exp{(-\chi^2/2)}$ and then binning them as a function of the parameter value. More detailed descriptions can be found in  \cite{kauffmann2003stellar} or \cite{salim2007uv}.

The measurements of star formation rate (SFR) are also taken from the MPA-JHU catalog, which are the median values of the total SFR PDFs. For SFGs with strong emission lines, SFRs within the SDSS fiber aperture are estimated from emission lines as described in \cite{brinchmann2004physical}, while the ones beyond the fiber are computed from the galaxy photometry, namely the similar photometric method used to derived $M_*$ described above, but ultraviolet photometry is involved to constrain SFR. For galaxies with weak emission lines (e.g., our PSB sample), SFRs are estimated via the photometric method using the integrated ultraviolet/optical photometry. We refer the reader to \cite{salim2007uv} for a detailed introduction.

\subsection{Star-forming and Quiescent Galaxies}

\begin{figure}[ht!]
\includegraphics[width=0.5\textwidth]{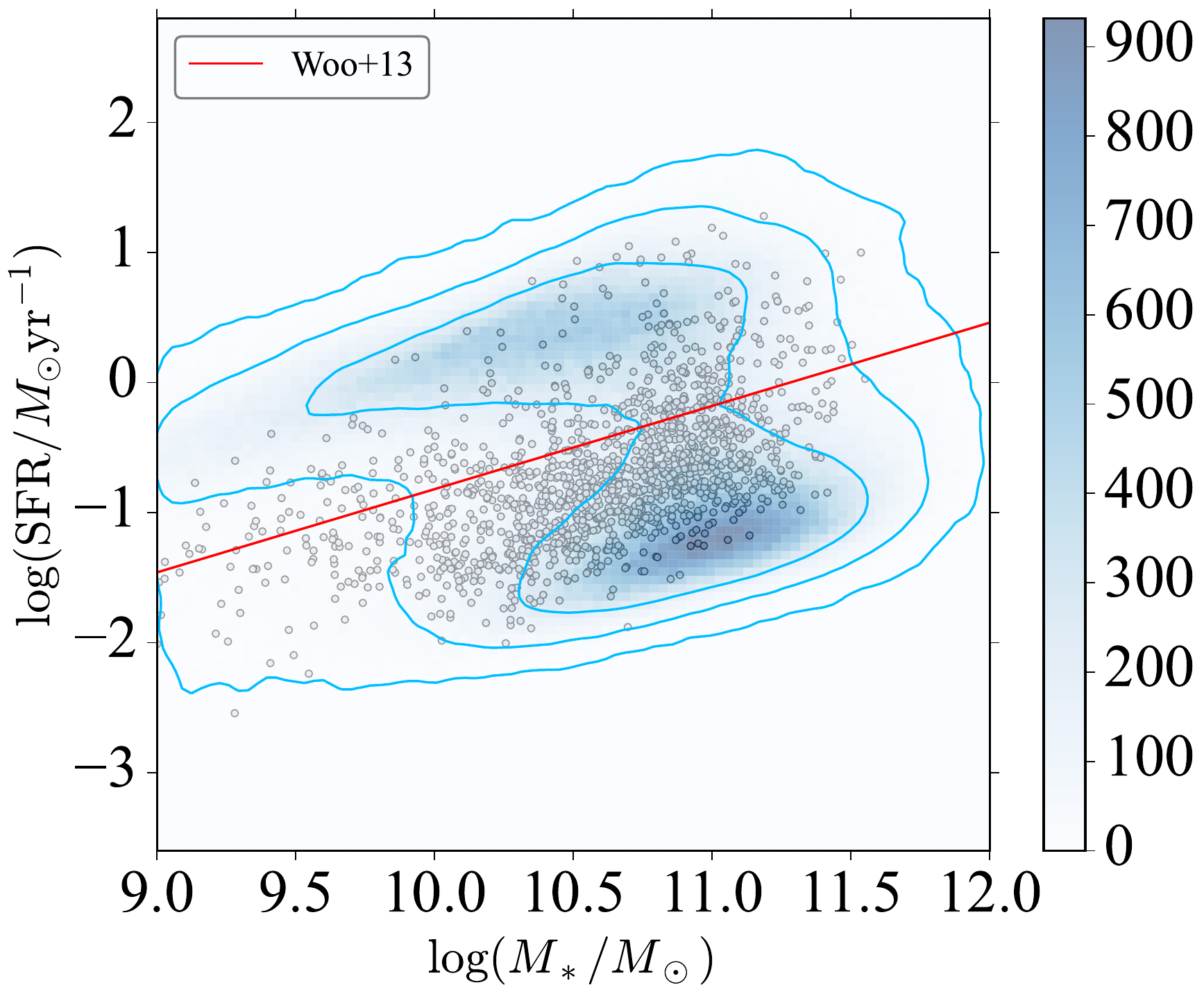}
\caption{SFR versus $M_*$ for our selected PSB candidates. The grep points indicate the distribution of individual PSBs. The color map shows the number density of all galaxies used in this work (e.g., Samples SFG, PSB, and QG), while the contours encompass 99.7\%, 95.5\%, and 68.3\% of them, respectively. The red solid line is the divided line for SFGs and QGs following \citet{woo2013dependence}.
\label{fig:dis}}
\end{figure}

To compare PSB galaxies with SFGs and QGs, we follow \citet{woo2013dependence} to divide galaxies into star-forming and quiescent galaxies based on the divided line of $\log \mathrm{SFR} (M_\odot~\mathrm{yr}^{-1}) = 0.64 \times \log M_*(M_\odot) -7.22$. 
313,427 SFGs and 327,700 QGs are selected and defined as Sample SFG and Sample QG, respectively. The distributions of SFGs, QGs, and PSBs on the $M_*$--SFR diagram are shown in Figure \ref{fig:dis}.
The red solid line is the divided line from \citet{woo2013dependence}, which clearly indicates the division between SFGs and QGs in our sample.  
It is not surprising that most of the PSBs are located near the divided line with a relatively large scatter, because this population is mainly known as galaxies in a transition phase of galaxy evolution for which the specific pathway, however, is still in debate \citep{pawlik2018origins,pawlik2019diverse}.

\subsection{Control Sample}
\label{subsec:control}

As the sample size of Sample PSB is much smaller than that of Sample SFG or QG, we generate two control samples to better compare the relation between SFGs, QGs, and PSB galaxies. We take Sample SFG or QG as a parent sample and create control sample with similar redshift and stellar mass. For each galaxy in Sample PSB, we search its nearest neighbors in the Sample SFG (or Sample QG) on the $M_*$--$z$ plane within a distance of 0.03 dex in $\log M_*(M_{\odot})$ and 0.005 in $z$. Each galaxy in the parent samples only enters the control sample once. 
If less than three galaxies are matched, we expand the $\Delta \log M_*(M_{\odot})$ and $\Delta z$ tolerances by 0.01 dex and 0.001, respectively. 95.3\% of PSB galaxies can obtain sufficient matches in the first search.
In the end, each galaxy in Sample PSB has three matched SFGs and QGs with similar $M_*$ and redshift. Thus, the resulting Sample Control SFG and Sample Control QG both have a sample size of 4,437.

\section{Results}\label{sec:result}

\subsection{The $M_*$ Distribution of PSBs}\label{subsec:M_dist}

\begin{figure}[htb!]
\includegraphics[width=0.5\textwidth]{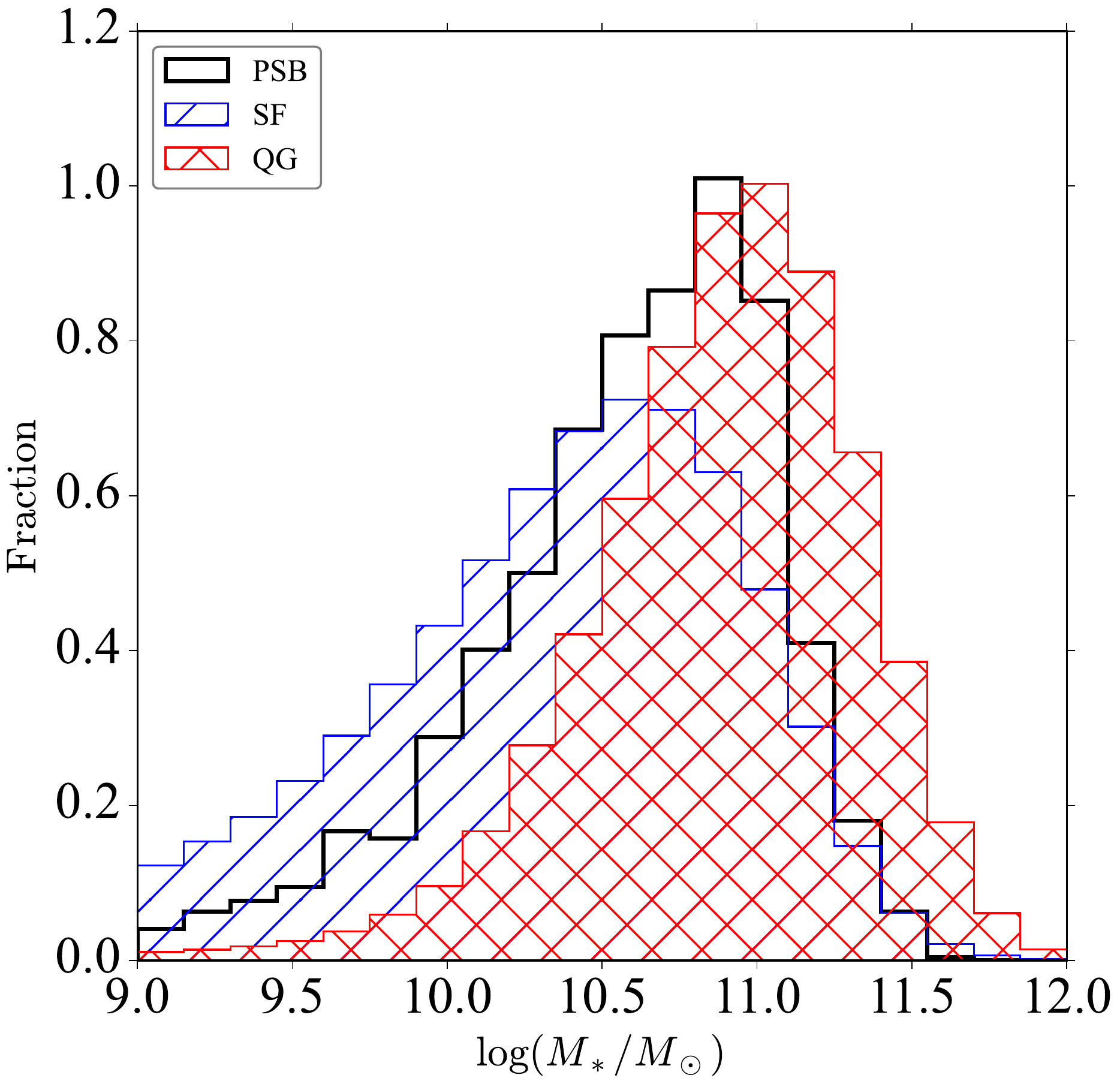}
\caption{ 
The distribution of stellar mass for Samples PSB (black), SFG (blue), and QG (red).
\label{fig:m_dis}}
\end{figure}

Firstly, we show the stellar mass distribution of PSB galaxies in Figure \ref{fig:m_dis}. Clearly, QGs tend to be more massive compared to the other two samples, while the median of $\log (M_*/M_{\odot})$ are 10.93, 10.66, and 10.47 for QGs, PSBs, and SFGs, respectively. \cite{maltby2018structure} reported a strong redshift evolution of the $M_*$ distribution for their photometrically selected PSB sample (i.e., $M_*\lesssim 10^{10}~\Msun$ for most of their PSBs at $0.5<z<1$ and $M_*\gtrsim 10^{10}~\Msun$ for those at $1<z<2$), suggesting less massive PSB galaxies in lower-redshift Universe. However, the majority (88.8\%) of our local PSBs have $M_*$ greater than $10^{10}~\Msun$. Such apparent conflict is due to the fact that our samples are selected from the MPA-JHU catalog released in the SDSS DR8, within which most of galaxies are from a flux-limited sample. The main galaxy spectroscopic survey of SDSS contributes most of galaxies in the MPA-JHU catalogs\footnote{\url{https://wwwmpa.mpa-garching.mpg.de/SDSS/DR7/Data/alldr72spectro-mpajhu.par}} and is sampled to a $r$-band magnitude limit of 17.77 mag \citep{strauss2002spectroscopic}, corresponding to a stellar mass of $M_*\gtrsim 2\times10^{10}~\Msun$ at the median redshift of Sample PSB ($z=0.13$) for a typical color of green valley galaxies (e.g., \citealt{schawinski2014green}) based on the \cite{bell2003optical} calibration. In other words, our PSB sample is naturally biased toward the massive end of low-redshift PSB galaxies, although less massive PSBs (i.e., $M_*\lesssim 10^{10}~\Msun$) are expected to be predominant in this population \citep{wild2016evolution,maltby2018structure}.

As the sample size of PSBs is small, we are unable to construct a mass complete sample to remove any potential bias arising from sample selection when comparing with SFGs and QGs. Instead, we will use control samples described in Section \ref{subsec:control} in the following analysis to avoid this problem.

\subsection{Size--Mass Relation}
\label{subsec:size_mass}

\begin{figure*}[htb!]
\includegraphics[width=0.5\textwidth]{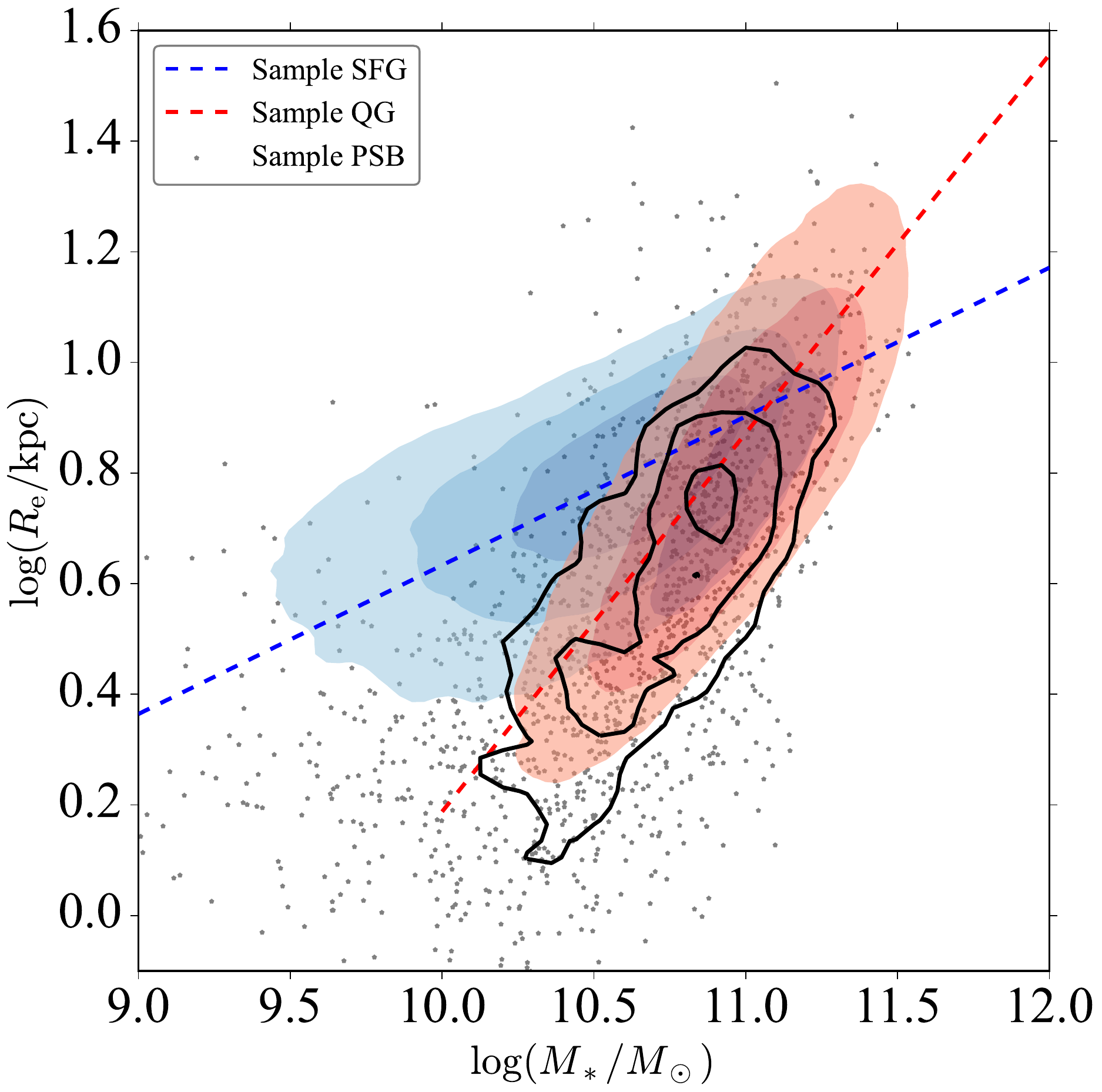}
\includegraphics[width=0.5\textwidth]{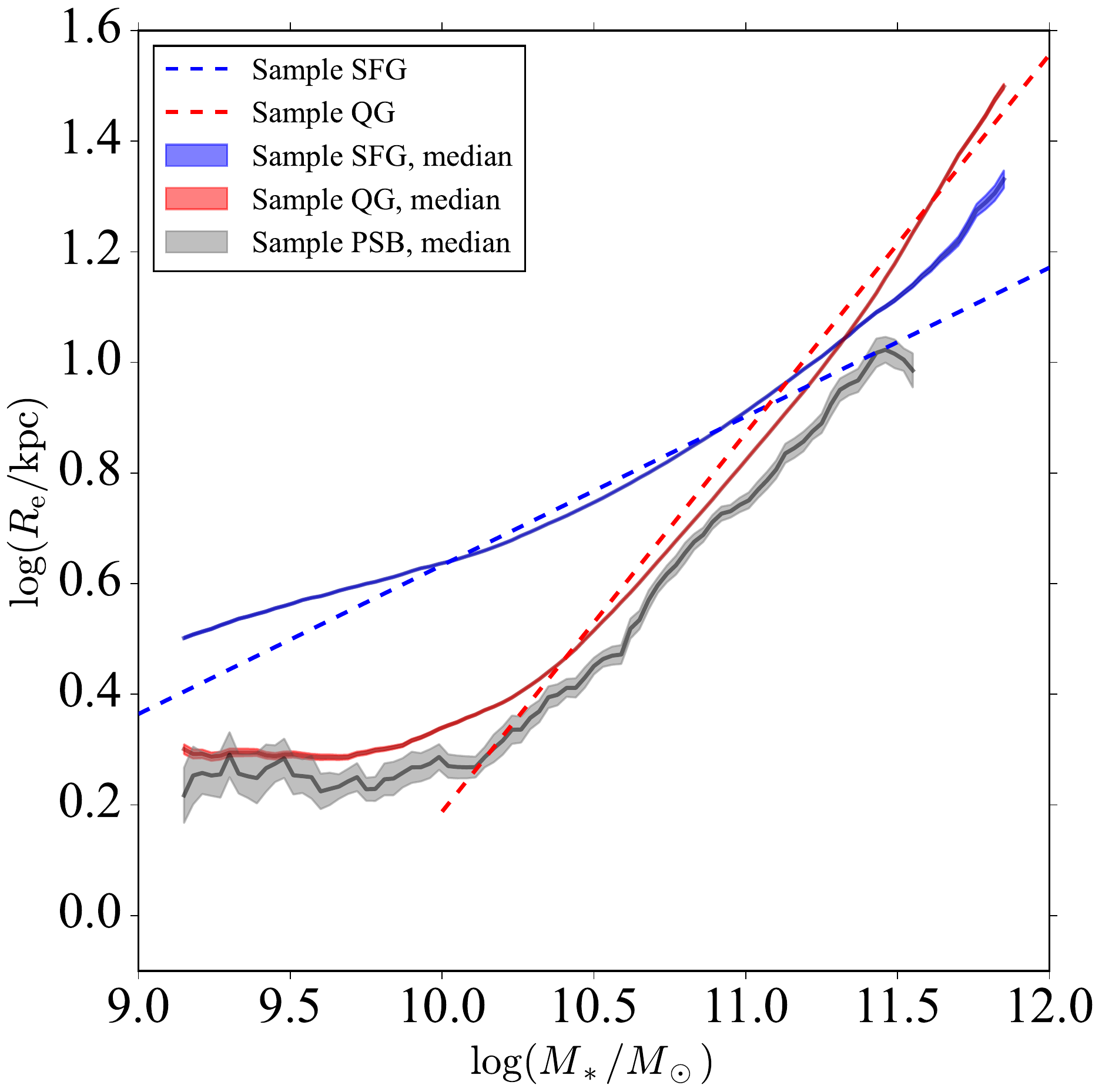}
\caption{Size--mass relation for our samples. Left: the overall distributions of SFGs (blue filled contours), QGs (red filled contours), and PSBs (black open contours) on the size--mass diagram. The contours enclose 25\%, 50\%, and 75\% of the corresponding samples, respectively. The blue (red) dashed line is the best-fit for the size--mass relation in formula of $R_{\mathrm{e}}/\mathrm{kpc} = \mathrm{A} \times m_{*}^{\alpha}$ ($m_*\equiv M_*/7\times 10^{10}~M_{\odot}$; \citealt{vanderwel20143dhst}) for SFGs (QGs). The grey points exhibit the distribution of individual PSBs.
Right: the running median of galaxy size as a function of $M_*$ for Samples SFG (blue), QG (red), and PSB (grey). The solid curves represent the running median values, while the shadow regions denote the corresponding 1-$\sigma$ uncertainties. Only bins with more than 10 galaxies are plotted. The dashed lines are the same as those in the left panel.
\label{fig:re-m}}
\end{figure*}

In this section, we show the typical size--mass relation for SFGs, QGs, and PSB galaxies. In the left panel of Figure \ref{fig:re-m}, we present the overall distributions of Samples SFG, QG, and PSB on the stellar size--mass diagram. To parameterize their overall trends, we follow \cite{vanderwel20143dhst} and fit the distributions with a formula of $R_{\mathrm{e}}/\mathrm{kpc} = \mathrm{A} \times m_{*}^{\alpha}$ ($m_*\equiv M_*/7\times 10^{10}~M_{\odot}$) in the $M_*$ ranges of $10^{9}~M_{\odot} < M_* < 10^{12}~M_{\odot}$ and $10^{10}~M_{\odot} < M_* < 10^{12}~M_{\odot}$ for SFGs and QGs, respectively. The best-fit results are also plotted in Figure \ref{fig:re-m}.

Although the PSB galaxies span a relatively large range in size at a fixed $M_*$, the number density contour of this population exhibits a large overlap with that of QGs and is only slightly lower than the best-fit size--mass relation of QGs,
suggesting a very similar size--mass relation between these two populations. On the other hand, the sizes of PSBs are much smaller than those of SFGs at a fixed $M_*$, while the differences become larger towards the low-mass end.

In order to clearly exhibit the difference in the size--mass relation between Samples PSB and QG, we calculate the running medians for the three samples at $10^{9}~\Msun < M_* < 10^{12}~\Msun$, which are shown in the right panel of Figure \ref{fig:re-m}. Only bins with more than 10 galaxies are considered and plotted. The 1-$\sigma$ uncertainties of the running medians are also computed via the bootstrapping method and denoted as shadow regions in the figure. The median size--mass relations of Samples SFG and QG have a cross at the most massive end (i.e., $M_*\sim 10^{11.0}~\Msun$), which is consistent with the finding in low-redshift Universe based on either deep field (e.g., \citealt{vanderwel20143dhst,mowla2019cosmosdash}) or wide field (e.g., \citealt{mosleh2013robustness,roy2018evolution}) surveys. Obviously, SFGs are found to have the largest galaxy size in most of the $M_*$ bins. As indicated by the relative position between the contours of PSBs and the best-fit size--mass relation of QGs, the median sizes of PSBs are smaller than those of QGs, suggesting that PSBs seem to be more compact than QGs. This result is further examined using the control samples in Section \ref{subsec:comp_control} since we have demonstrated that Sample PSB is not mass complete (see Section \ref{subsec:M_dist}).

\subsection{Comparison with Control Samples}
\label{subsec:comp_control}

\begin{figure}[htb!]
\includegraphics[width=0.5\textwidth]{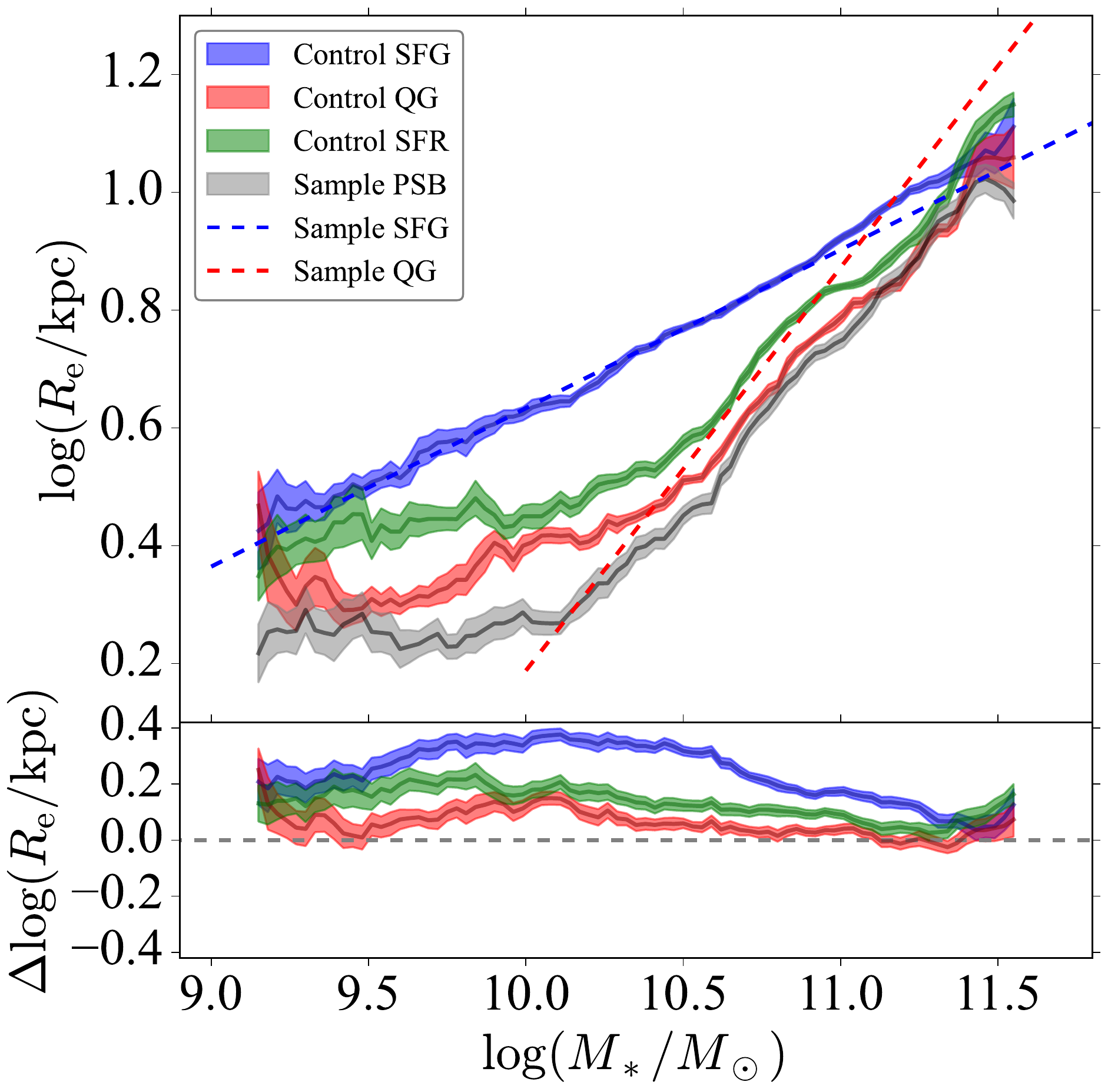}
\caption{Size--mass relations for Sample PSB and three control samples. In the top panel, the blue, red, green, and black solid curves indicate the running medians of Samples Control SFG, Control QG, Control SFR (see text for more details), and PSB, respectively. The shadow regions around each median curve denote the corresponding 1-$\sigma$ uncertainties. The blue and red dashed lines are the same as those in the left panel of Figure \ref{fig:re-m} (i.e., the best-fit relations for Samples SFG and QG, respectively). In the bottom panel, the differences in the running medians between Sample PSB and the three control samples are presented, together with the 1-$\sigma$ uncertainties denoted as shadow regions.}
\label{fig:control_sample}
\end{figure}

As mentioned earlier, we attempt to utilize the control samples to remove any potential bias due to the incompleteness of sample construction. Here, we show the size--mass relation in the $M_*$ range of $10^9~\Msun < M_* < 10^{12}~\Msun$ for Sample PSB, Control SFG, and Control QG in Figure \ref{fig:control_sample}. The running medians of galaxy size of these samples and the corresponding uncertainties via the bootstrapping method are calculated.

Comparing with the best-fit size--mass relation of Sample QG, although the median relation of Control QG shows a small decrease at $M_*\gtrsim 10^{10.5}~\Msun$, the median galaxy sizes of PSBs are still smaller than those of QGs in most of the $M_*$ we explored. Such differences between PSBs and QGs become more significant towards the low-mass end. Thus we conclude that PSBs tend to have a more compact morphology compared to Control QG at $M_*\lesssim 10^{11.0}~\Msun$. On the other hand, the median size--mass relation of Control SFG only has small variations around the median relation of its parent sample, leading to a much larger galaxy size of SFG population than that of Sample PSB after controlling redshift and $M_*$. At $M_*\sim 10^{11.5}~\Msun$, the median size--mass relations of the three populations are consistent with each other within the uncertainties.

We summarize the median values of $R_{\rm e}$ and the corresponding 1-$\sigma$ uncertainties in bins of $M_*$ for Sample PSB, Control SFG, and Control QG, together with the numbers of galaxies within each bin ($N$), in Table \ref{tab:1}. To quantify the strength of the differences, we perform Kolmogorov-Smirnov (KS) tests for the differences in $R_{\rm e}$ between Sample PSB and other two control samples. A KS test $p$-value of 0.05 means that there is 95 per cent confidence that we can reject the null hypothesis that the two samples are drawn from the same distribution. The one-by-one $R_{\rm e}$ differences in bins of $M_*$ and the resulting $p$-values from KS tests are shown in Table \ref{tab:2}. The galaxy sizes of PSBs are estimated be on average $\sim 13\%$ smaller than those of QGs, especially at $10^{9.5}~\Msun \lesssim M_*\lesssim 10^{10.5}~\Msun$ where PSBs can be on average $\sim 19\%$ smaller than QGs. Obviously, the size differences between QGs and PSBs are significant at $10^{9.6}~\Msun \lesssim M_*\lesssim 10^{11.1}~\Msun$, with all the $p$-values below 0.05.

As shown in Figure \ref{fig:dis}, the distribution of PSB galaxies in the $M_*$--SFR plane biases toward the QG side. One might wonder whether this distribution bias has any effect on the size distribution of PSBs. Observationally, PSBs in the color--mass (or magnitude) plane tend to reside in the GV region (e.g., \citealt{wong2012galaxy,greene2021refining}), within which galaxies are believed to be in a transitional phase and have larger size compared to QGs \citep{salim2014green,gu2018morphological,suess2021dissecting}. Therefore, we would not expect that the distribution bias in $M_*$--SFR plane has any contribution to the compact morphology of PSBs. However, to totally unveil the role of this bias, we further construct a three-parameter control sample using the method described in Section \ref{subsec:control} but control SFR at the same time (i.e., a $M_*$--SFR--$z$ control sample, defined as Sample Control SFR hereafter). The median size--mass relation for this new control sample is given in Table \ref{tab:1} and denoted by green curve in Figure \ref{fig:control_sample}, together with its 1-$\sigma$ uncertainty.

As expected, Sample Control SFR has an intermediate galaxy size between SFGs and QGs, which is significantly larger than that of Sample PSB. Namely, the removal of the distribution bias in the $M_*$--SFR plane, on the contrary, reinforces the result of an extreme compact morphology for PSBs described earlier.

In the bottom panel of Figure \ref{fig:control_sample}, we show the differences of the running medians between Sample PSB and the control samples. The results from the top panel are more obvious in this plot. 
In short, the sizes of PSBs are smaller compared to QGs after controlling $M_*$ and redshift, and the differences between them become larger and significant towards the low-mass end.

\begin{deluxetable}{cccc}
\tablecaption{Median sizes for Samples PSB, Control SFG, Control QG, and Control SFR.}
\label{tab:1}
\tablewidth{0pt}
\tablehead{
\colhead{Sample} & \colhead{$\log(M_*/\Msun)$} & \colhead{$\log (R_{\rm e}/\rm kpc)$} & \colhead{$N$}
}

\startdata
Sample PSB& 9.0-9.3&  0.22 $\pm$ 0.04& 23\\
& 9.3-9.6&  0.27 $\pm$ 0.03& 38\\
& 9.6-9.9&  0.23 $\pm$ 0.02& 72\\
& 9.9-10.2&  0.27 $\pm$ 0.02& 153\\
& 10.2-10.5&  0.39 $\pm$ 0.02& 263\\
& 10.5-10.8&  0.53 $\pm$ 0.02& 371\\
& 10.8-11.1&  0.73 $\pm$ 0.01& 413\\
& 11.1-11.4&  0.89 $\pm$ 0.02& 131\\
& 11.4-11.7&  0.99 $\pm$ 0.03& 15\\
Control SFG& 9.0-9.3&  0.43 $\pm$ 0.07& 67\\
& 9.3-9.6&  0.50 $\pm$ 0.02& 113\\
& 9.6-9.9&  0.58 $\pm$ 0.02& 219\\
& 9.9-10.2&  0.64 $\pm$ 0.01& 460\\
& 10.2-10.5&  0.73 $\pm$ 0.01& 786\\
& 10.5-10.8&  0.81 $\pm$ 0.01& 1114\\
& 10.8-11.1&  0.90 $\pm$ 0.01& 1239\\
& 11.1-11.4&  1.01 $\pm$ 0.01& 390\\
& 11.4-11.7&  1.11 $\pm$ 0.05& 49\\
Control QG& 9.0-9.3&  0.47 $\pm$ 0.06& 67\\
& 9.3-9.6&  0.29 $\pm$ 0.02& 114\\
& 9.6-9.9&  0.33 $\pm$ 0.02& 218\\
& 9.9-10.2&  0.42 $\pm$ 0.01& 457\\
& 10.2-10.5&  0.45 $\pm$ 0.01& 784\\
& 10.5-10.8&  0.58 $\pm$ 0.01& 1116\\
& 10.8-11.1&  0.77 $\pm$ 0.01& 1236\\
& 11.1-11.4&  0.90 $\pm$ 0.02& 395\\
& 11.4-11.7&  1.06 $\pm$ 0.05& 50\\
Control SFR& 9.0-9.3&  0.35 $\pm$ 0.04& 65\\
& 9.3-9.6&  0.45 $\pm$ 0.05& 110\\
& 9.6-9.9&  0.45 $\pm$ 0.02& 209\\
& 9.9-10.2&  0.45 $\pm$ 0.02& 426\\
& 10.2-10.5&  0.53 $\pm$ 0.01& 806\\
& 10.5-10.8&  0.64 $\pm$ 0.01& 1094\\
& 10.8-11.1&  0.83 $\pm$ 0.01& 1236\\
& 11.1-11.4&  0.93 $\pm$ 0.02& 396\\
& 11.4-11.7&  1.15 $\pm$ 0.02& 44\\
\enddata
\end{deluxetable}

\begin{deluxetable*}{cccccc}
\tablecaption{Differences in galaxy sizes and concentrations and the corresponding $p$-values from KS tests between PSBs and two control samples.}
\label{tab:2}
\tablewidth{0pt}
\tablehead{
\colhead{Sample} & \colhead{$\log(M_*/\Msun)$} & \colhead{$\Delta \log (R_e/\rm kpc)$} & \colhead{$p$-value ($R_{\rm e}$)} & \colhead{$\Delta C$} & \colhead{$p$-value ($C$)}
}

\startdata
SF - PSB& 9.0-9.3&  0.21 $\pm$ 0.08& 0.02& -0.41 $\pm$ 0.06& $<$0.01\\
& 9.3-9.6&  0.23 $\pm$ 0.04& $<$0.01& -0.37 $\pm$ 0.06& $<$0.01\\
& 9.6-9.9&  0.35 $\pm$ 0.03& $<$0.01& -0.40 $\pm$ 0.08& $<$0.01\\
& 9.9-10.2&  0.37 $\pm$ 0.02& $<$0.01& -0.44 $\pm$ 0.02& $<$0.01\\
& 10.2-10.5&  0.34 $\pm$ 0.02& $<$0.01& -0.56 $\pm$ 0.02& $<$0.01\\
& 10.5-10.8&  0.27 $\pm$ 0.02& $<$0.01& -0.59 $\pm$ 0.02& $<$0.01\\
& 10.8-11.1&  0.17 $\pm$ 0.01& $<$0.01& -0.61 $\pm$ 0.02& $<$0.01\\
& 11.1-11.4&  0.12 $\pm$ 0.02& $<$0.01& -0.52 $\pm$ 0.02& $<$0.01\\
& 11.4-11.7&  0.13 $\pm$ 0.06& 0.04& -0.45 $\pm$ 0.09& $<$0.01\\
QG - PSB& 9.0-9.3&  0.25 $\pm$ 0.08& 0.01& -0.29 $\pm$ 0.07& $<$0.01\\
& 9.3-9.6&  0.02 $\pm$ 0.04& 0.46& -0.07 $\pm$ 0.06& 0.12\\
& 9.6-9.9&  0.10 $\pm$ 0.03& $<$0.01& -0.11 $\pm$ 0.09& $<$0.01\\
& 9.9-10.2&  0.15 $\pm$ 0.02& $<$0.01& -0.07 $\pm$ 0.03& $<$0.01\\
& 10.2-10.5&  0.05 $\pm$ 0.02& $<$0.01& -0.09 $\pm$ 0.02& $<$0.01\\
& 10.5-10.8&  0.05 $\pm$ 0.02& 0.02& -0.09 $\pm$ 0.02& $<$0.01\\
& 10.8-11.1&  0.04 $\pm$ 0.01& $<$0.01& -0.05 $\pm$ 0.02& $<$0.01\\
& 11.1-11.4&  0.01 $\pm$ 0.03& 0.13& 0.01 $\pm$ 0.02& 0.07\\
& 11.4-11.7&  0.07 $\pm$ 0.06& 0.05& 0.06 $\pm$ 0.08& 0.28\\
\enddata

\end{deluxetable*}

\subsection{Differences in Concentration and S\'ersic Index}

\begin{figure}[htb!]
\centering
\includegraphics[width=0.5\textwidth]{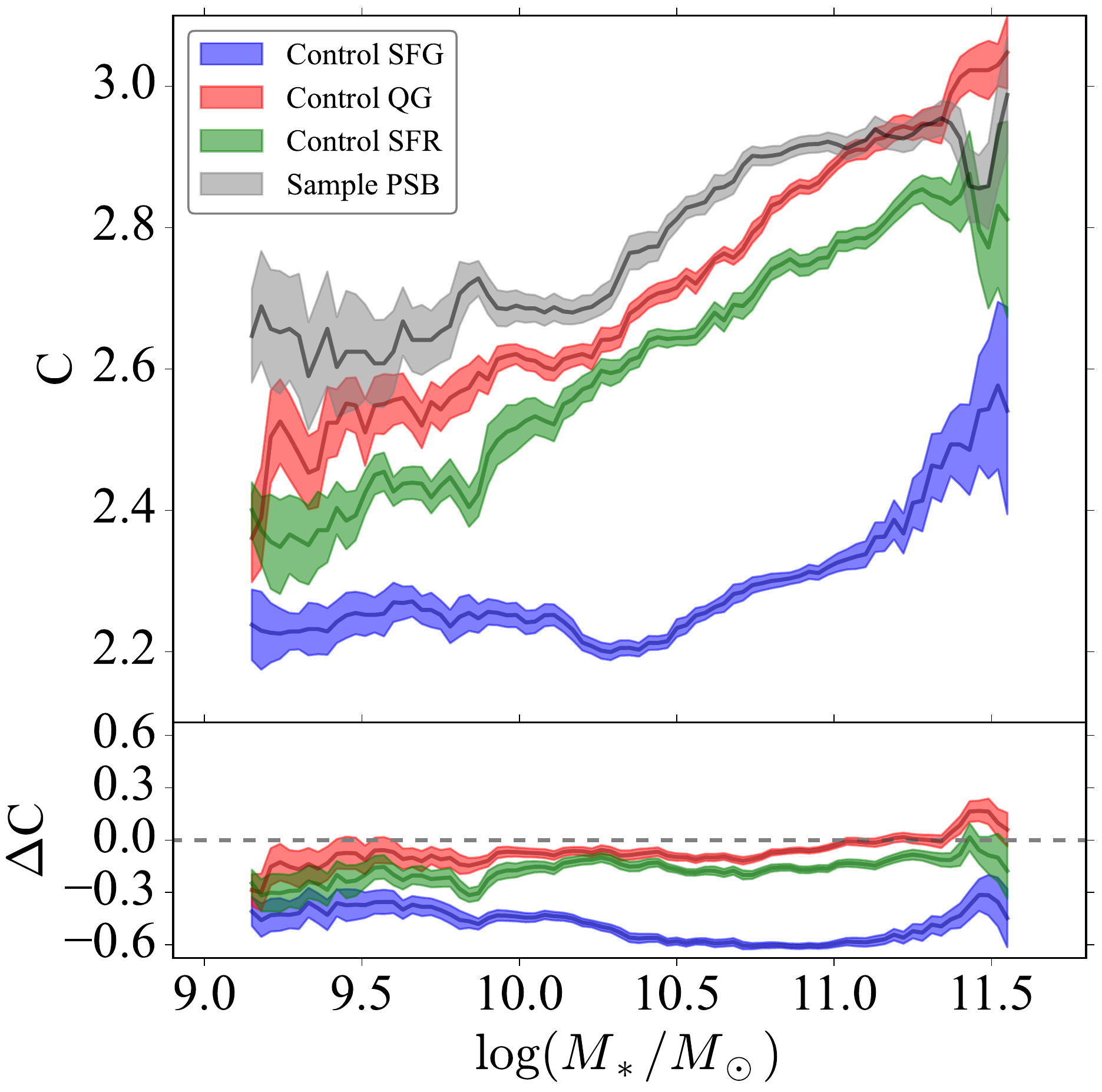}
\caption{Comparisons of concentration $R_{\rm 90,r}/R_{\rm 50,r}$ between Sample PSB and three control samples. The symbols are the same as in Figure \ref{fig:control_sample}.}
\label{fig:concentration}
\end{figure}

Besides the effective radius $R_{\rm e}$, the more compact morphology of PSB galaxies compared to QGs also can be traced by the concentration $C$ or the best-fit S\'ersic index $n$. We show the comparisons of the concentration between Sample PSB and three control samples in Figure \ref{fig:concentration}. Here, the concentration is defined as $C=R_{\rm 90,r}/R_{\rm 50,r}$ \citep{graham2005total}, where $R_{\rm 50,r}$ and $R_{\rm 90,r}$ are radii containing 50\% and 90\% of the Petrosian fluxes in $r$-band, respectively. Choosing the concentration defined by Petrosian fluxes (rather than the ones from the single S\'ersic fit where our adopted $R_{\rm e}$ is derived) ensures an independent verification of our results.

It is clear that PSBs have the largest $C$ at $M_{*}\lesssim 10^{11.0}~\Msun$ compared to all of the three control samples. The differences in $C$ in bins of $M_*$ are also given in Table \ref{tab:2}, together with the corresponding $p$-values of KS tests. The concentrations of PSBs are found to be larger than those of Control QGs by 0.07 on average, while the KS tests reveal that such differences are significant at a confidence level of 99\% at $10^{9.6}~\Msun \lesssim M_* \lesssim 10^{11.1}~\Msun$. These results are in good agreement with our finding based on the galaxy sizes.

Similar comparisons using the best-fit S\'ersic index $n$ extracted from the \cite{meert2015catalogue} catalog are also examined. Due to the strong correlation between the concentrations and the S\'ersic index $n$ \citep{graham2005total}, we will not show the detailed results for $n$ and just report that the index $n$ of PSBs is larger than that of Control QG by 1.01 on average and the differences are significant at a confidence level of 99\% at $10^{9.6}~\Msun \lesssim M_* \lesssim 10^{11.1}~\Msun$.

\section{Discussion}\label{sec:discussion}

In this work, we investigate the size--mass relation for PSB galaxies in the local Universe and compare it with that of SFGs and QGs. Our results reveal a more compact morphology for PSBs compared to QGs with similar $M_*$ for $M_*\lesssim 10^{11.0}~\Msun$, especially at $10^{9.5}~\Msun \lesssim M_*\lesssim 10^{10.5}~\Msun$ where the difference in galaxy size becomes larger and significant, indicating that similar morphological feature observed in higher-redshift PSBs \citep{almaini2017massive,wu2018fast} still exists in the local Universe. In this section, we discuss possible implication of these results.

\subsection{Different Evolution Pathways of PSBs}
\label{subsec:diff-pathways}

Previous works focused on the intermediate- or high-redshift (i.e., $z\gtrsim 0.5$) PSBs found that massive PSB galaxies (i.e., $M_*\gtrsim 10^{10}~\Msun$) have smaller or similar galaxy size in comparison with QGs at the same epoch (e.g., \citealt{belli2015stellar,yano2016relation,almaini2017massive,wu2018fast,wu2020colors}), supporting a fast quenching scenario in which galaxy structure is dramatically changed via a wet compaction before/during the quenching of star formation (e.g., \citealt{dekel2009cold,tacchella2016evolution}). However, few of these works reach the mass range of $M_*\lesssim 10^{10}~\Msun$.

On the other hand, observational studies about the local PSBs highlighted the diversity of this population and claimed various evolutionary pathways of PSBs including the fast quenching with significant structural changes and cyclic evolution without substantial morphological transformation \citep{pawlik2018origins}. Based on the data products from cosmological simulations, \cite{pawlik2019diverse} further presented detailed descriptions of four typical examples of PSB galaxies, three of which are able to produce galaxy sizes as compact as QGs at a fixed $M_*$ in their PSB phase. Since our results reveal that the morphology of PSBs resemble QGs rather than SFGs, we will attribute the dominated nature of our PSB sample to one of these compact PSBs in this section, while the last example that produces more disk-like, extended PSBs will be discussed in Section \ref{subsec:environment}.

The three compact PSB prototypes represent three evolutionary pathways that all contain a distinct short-lived starburst and thus are able to exhibit observational spectral features of PSB galaxies, including a blue-to-red fast quenching, a blue-to-blue cycle (i.e., a SFG passes through a PSB phase after which the galaxy still maintains low level star formation), and a red-to-red rejuvenation (i.e., a QG enters a PSB phase and finally returns to quiescent phase). The evident starbursts of the former two pathways are induced by a major merger, while the enhancement in SFR of the last one is due to a minor merger.

PSB galaxies in these channels are all able to have galaxy sizes comparable with those of QGs at a fixed $M_*$ according to the similar concentrations between these two populations\footnote{Both the blue-to-red quenching and the blue cycle undergo significant merger-induced morphological changes and the growth of a central spheroidal component, while the red cycle happens to a QG that already show spheroidal morphology.} in simulations \citep{pawlik2019diverse} and thus might have contributions to our PSB sample in observations. However, PSBs in the blue cycle are expected to have measurable H$\alpha$ emission due to their ongoing (low level) star formation activities, which are already rejected by our selection criterion of no-H$\alpha$ emission line described in Section \ref{subsec:selection}. On the one hand, we also note that the time that the galaxy spent in the PSB phase of the red cycle is much smaller than that of the blue-to-red quenching (as shown in the Figure 3 of \citealt{pawlik2019diverse}), implying a very small possibility to observe a PSB in the red cycle (2\% of PSBs identified in their simulations). On the other hand, simulations predicted little morphological change during the red cycle, in contrast to the smaller galaxy size we observed for PSBs compared to that of QGs. Meanwhile, in simulations the blue-to-red channel displays a decline in concentration index after the burst, indicating a more compact morphology at the beginning of the PSB phase. The gradual decrease of the concentration index during the PSB phase directly reflects a slight increase of the (light-weighted) galaxy size as the galaxy ages and fades down and finally reaches the quiescent phase with a size similar to typical QGs. Such picture drawn from the simulations is in good agreement with our observations.

Therefore, the blue cycle should have no contribution to our sample by selection, while the red cycle (if have) should be not the main mechanism for our PSBs. In term of the evolution channel of PSBs, we believe that the blue-to-red fast quenching following a short-lived starburst (might be induced by major merger) should be the dominated one.

\subsection{Hint of Environmental Effect}
\label{subsec:environment}

Besides the three evolution channels with a short-lived burst of star formation, \cite{pawlik2019diverse} also claimed that about half of galaxies with PSB spectral features in their simulations have not underwent an evident merger-driven starburst event. This population is represented by the last example in \cite{pawlik2019diverse} for which the PSB features are caused by a rapid quenching of its continuous star formation after falling into a massive halo (galaxy cluster), i.e., an environmental effect.

Recent PSB studies about galaxy environment highlighted a nonnegligible effect of environment in the evolution of PSB galaxies \citep{poggianti2017gasp,paccagnella2019strong}, especially for low-$M_*$ PSB at $M_*\lesssim 10^{10}~\Msun$ \citep{socolovsky2018enhancement,wilkinson2021starburst}. Based on photometrically selected PSBs at $0.5<z<1$, \cite{socolovsky2018enhancement} found an excess of low-$M_*$ PSB galaxies in clusters compared to less dense environments, while \cite{wilkinson2021starburst} reported that low-$M_*$ PSBs prefer to reside in cluster-like environments. These studies suggested that fast quenching processes driven by dense environments might be one of the important mechanisms for low-$M_*$ PSB in the local and intermediate-redshift Universe, as predicted by the simulations of \cite{pawlik2019diverse}.

\begin{figure}[htb!]
\includegraphics[width=0.5\textwidth]{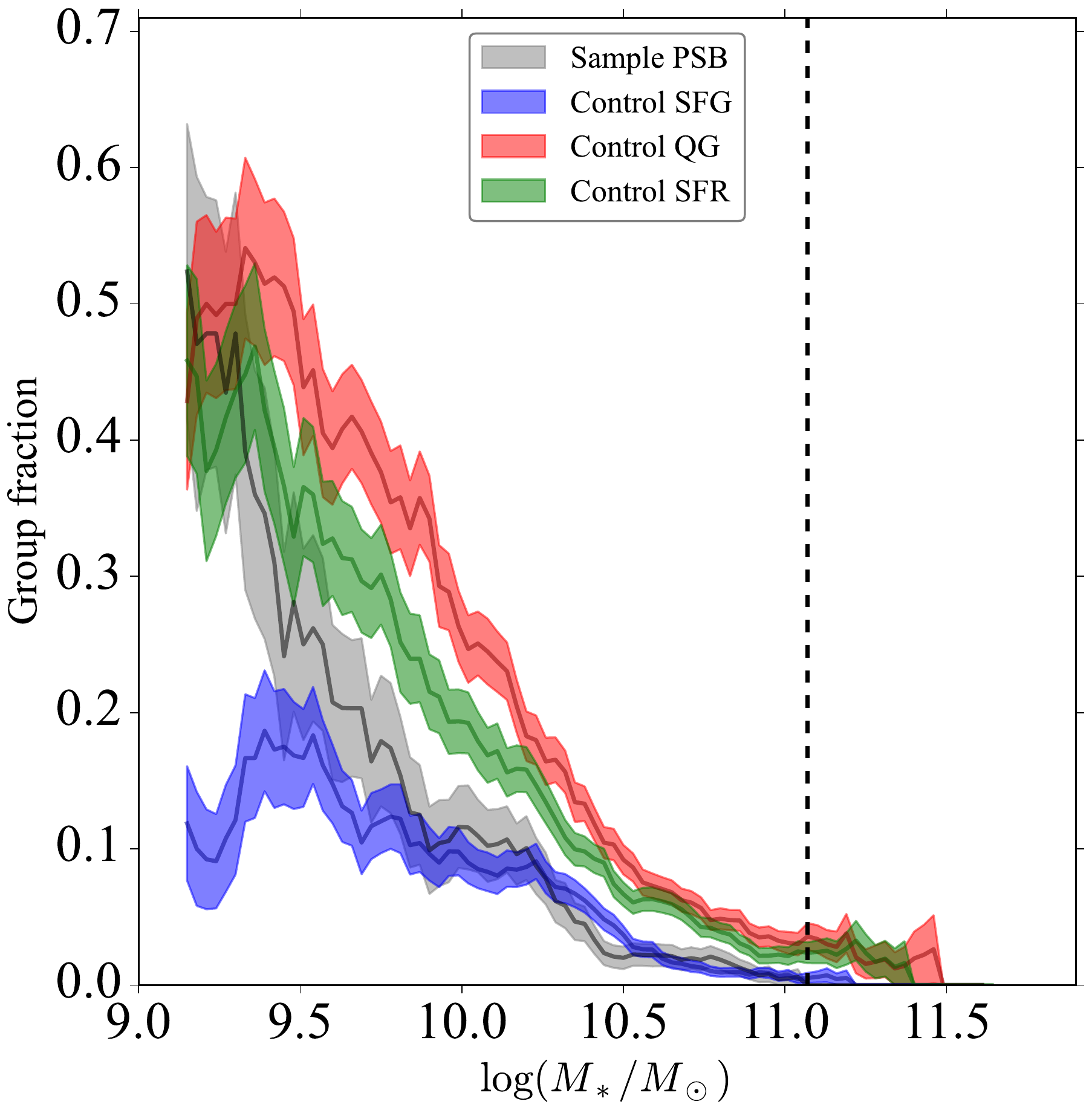}
\caption{Group fractions as a function of stellar mass for Sample PSB and three control samples. The black dashed line marks the mass threshold of $M_*=10^{11.07}~\Msun$ above which the fraction of the matched PSBs with the groups/clusters catalogs becomes lower than 50\%. Symbols are the same as those in Figure \ref{fig:control_sample}.}
\label{fig:f_group}
\end{figure}

\begin{figure*}[htb!]
\includegraphics[width=0.494\textwidth]{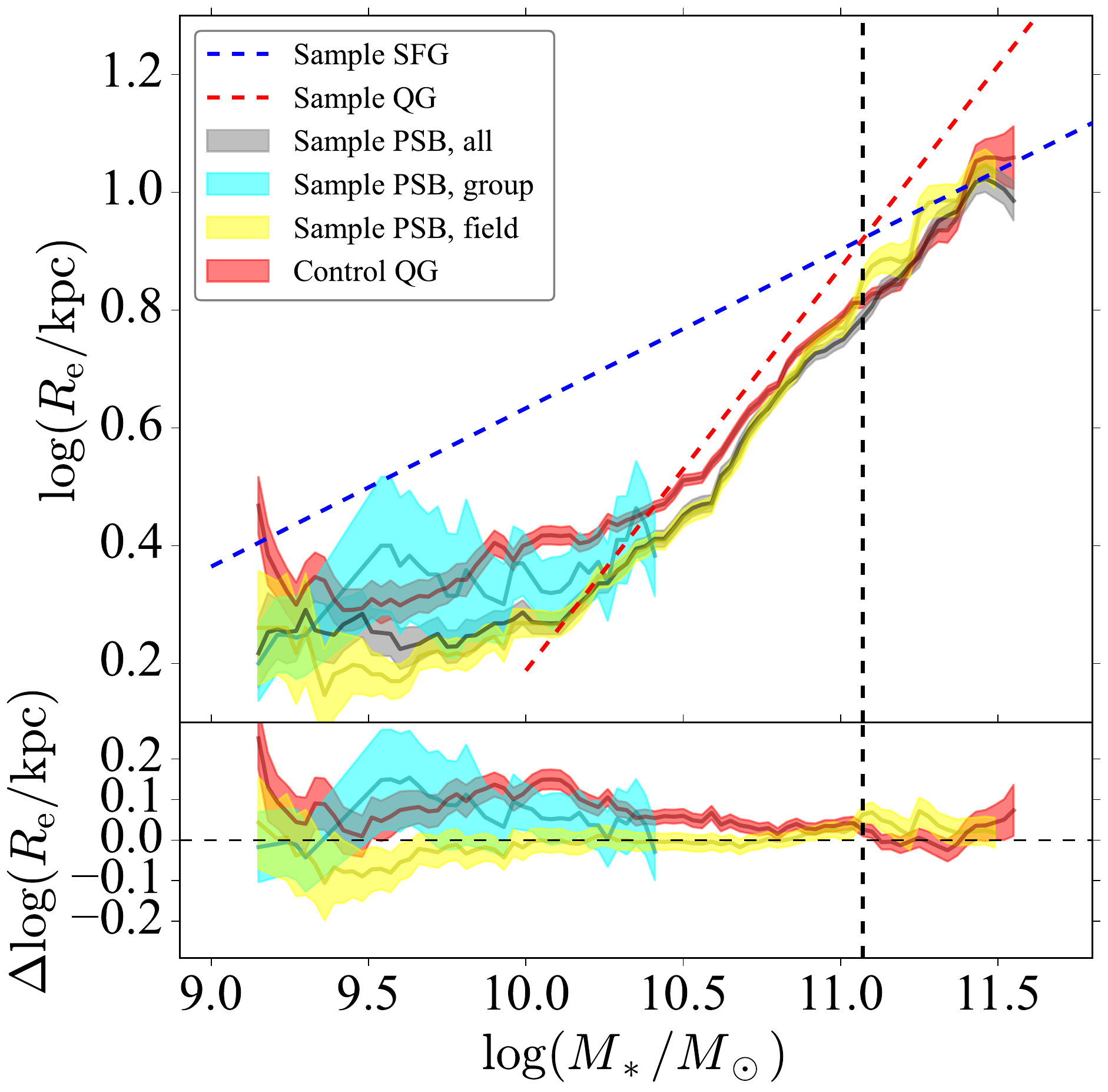}
\includegraphics[width=0.506\textwidth]{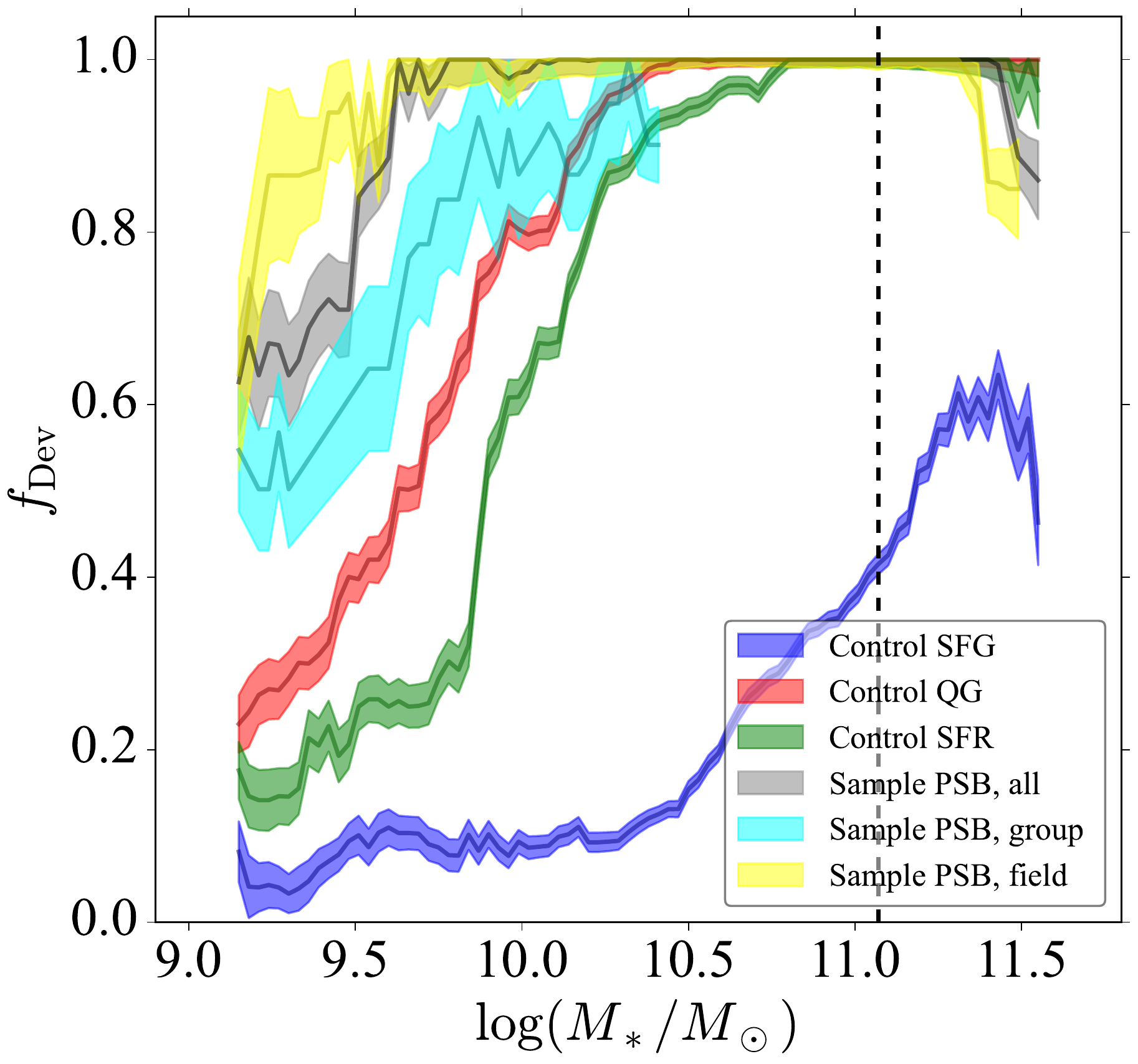}
\caption{Size--mass relations (left) and $f_{\mathrm{DeV}}$--mass relations for PSBs in different environments. The cyan (yellow) solid curves denote the median values of the corresponding parameters for PSBs in group (field), while the shadow regions are the 1-$\sigma$ uncertainties. The black dashed line in each panel marks the mass threshold of $M_*=10^{11.07}~\Msun$ above which the fraction of the matched PSBs with the groups/clusters catalogs becomes lower than 50\%. Other symbols are the same as those in Figure \ref{fig:control_sample}.}
\label{fig:psb_env}
\end{figure*}

To examine whether there is any environmental effect in our local PSB sample, we cross-match our samples with the catalogs of groups and clusters from \cite{tempel2014flux} to identify PSBs resided in high density environments, and thus calculate the group fraction ($f_{\mathrm{group}}$) that describes the fraction of PSBs resided in a galaxy group within each $M_*$ bin. Here, we define ``galaxy groups'' as those with number of members $N_{\mathrm{member}}\geq 10$. Similar criteria are also adopted by other studies, e.g., \cite{kim2020compact}. We note that the final matched PSBs account for only  76\% of Sample PSB due to two reasons: (1) a small fraction of sky sampled by our PSBs is not covered by the \cite{tempel2014flux} catalogs, and (2) these catalogs only contain galaxies out to $z=0.2$, while PSBs at $z>0.2$ account for about 15\% of our sample, which is also the most massive ones (i.e., $M_*\gtrsim 10^{11}~\Msun$). Figure \ref{fig:f_group} shows the group fraction as a function of $M_*$ for Sample PSB, as well as for the three control samples for comparison. The black dashed line marks the mass threshold of $M_*=10^{11.07}~\Msun$ above which the fraction of the matched PSBs with the groups/clusters catalogs becomes lower than 50\%, and this highest-$M_*$ regime will be ignored in the following discussion about the environmental effect.

Clearly, a general increase of the group fraction towards the low-$M_*$ end is observed for all the samples. Sample Control SFR has an intermediate $f_{\mathrm{group}}$ between Samples Control SFG and QG, similar to their distributions in the $M_*$--SFR plane. Meanwhile, the $f_{\mathrm{group}}$ fraction of Sample PSB at $M_*\gtrsim 10^{10}~\Msun$ is small (i.e., $f_{\mathrm{group}}\lesssim 10\%$), suggesting that most of massive PSBs are field galaxies and environment-related rapid quenching processes (if have) have little or no contribution to PSBs at the high-$M_*$ end. It is noteworthy that the local environment of these massive PSBs resembles that of Sample Control SFG rather than those of Samples Control SFR and QG, which is consistent with a picture in which a SFG undergoes a merger-induced starburst that followed by a rapid quenching and enters a PSB phase. At $M_*\lesssim 10^{10}~\Msun$, $f_{\mathrm{group}}$ of PSBs increases steeply towards the low-$M_*$ end, while its differences from those of Samples Control SFR and QG become smaller. Local PSBs at $M_*\lesssim 10^{10}~\Msun$ are more clustered than their high-$M_*$ analogues and SFGs with similar $M_*$. These trends are generally consistent with those reported in \cite{wilkinson2021starburst} for PSBs at $0.5<z<1.0$, implying a possible role of environment in the evolution of low-$M_*$ PSBs.

To further explore how environments impact the galaxy sizes of PSBs, we divide Sample PSB into two subsamples according to their environments and our aforementioned group definition. Namely, we define PSBs resided in a galaxy group with $N_{\mathrm{member}}\geq 10$ as group PSBs (i.e., Sample PSB, group), while the remaining isolated PSBs or those resided in a galaxy group with $N_{\mathrm{member}}<10$ are field PSBs (i.e., Sample PSB, field). The influence of adopting a different $N_{\mathrm{member}}$ criterion will be discussed in Section \ref{subsec:group_criterion}. In the left panel of Figure \ref{fig:psb_env}, we show the median mass--size relations for group and field PSBs. The relation for field PSBs is almost the same as that of the whole Sample PSB except at the lowest-$M_*$ end where field PSBs tend to be slightly smaller. The median mass--size relation shows a large dispersion, but it is still evident that the sizes of group PSBs tend to be larger compared to field PSBs. The $R_{\rm e}$ of group PSBs are $\sim 10\%$ larger on average than those of filed PSBs at $M_*\lesssim 10^{10}~\Msun$. However, the KS test in the same $M_*$ range only returns a $p$-value of 0.3, indicating that such differences are not significant enough, mainly due to the small number of group PSBs.

The environment-induced fast quenching process described in \cite{pawlik2019diverse} expects little morphological change and thus results in a disk-like PSB if the progenitor is also a star-forming disk before falling into the massive halo. As a simple check, we also use the flux fraction of the de Vaucouleurs component ($f_{\mathrm{DeV}}$), which quantifies the flux ratio of the best-fit de Vaucouleurs component to the total flux when an exponential and a de Vaucouleurs models are combined to fit the galaxy image (see the online algorithm description\footnote{\url{https://www.sdss.org/dr12/algorithms/magnitudes/\#cmodel}} for more details), in $r$-band extracted from the SDSS database to roughly evaluate the flux contribution of the central spheroidal component to the whole galaxy and whether a disk component is necessary. The $f_{\mathrm{DeV}}=1$ means that the galaxy is well described by a pure de Vaucouleurs profile, i.e., a compact spheroidal morphology without any disk component. The $f_{\mathrm{DeV}}$ as a function of $M_*$ for group and field PSBs, as well as control samples, are presented in the right panel of Figure \ref{fig:psb_env}.

Again, the trends of $f_{\mathrm{DeV}}$ for Samples Control SFG, SFR, and QG are in good agreement with our expectation in which Sample Control SFR still exhibits its intermediate nature. As analogues of these control samples in the $z$--$M_*$ or $z$--$M_*$--SFR parameter space, Sample PSB, however, shows a higher $f_{\mathrm{DeV}}$ (i.e., a much stronger spheroidal component), reinforcing our results demonstrated in Section \ref{sec:result}. Disk-like components are only necessary for PSBs at $M_*\lesssim 10^{10}~\Msun$. Moreover, $f_{\mathrm{DeV}}$ of field PSBs is comparable to that of the whole PSB sample and larger than that of group PSBs, especially at the lowest-$M_*$ end. A KS test between the $f_{\mathrm{DeV}}$ distributions of group and field PSBs at $M_*\lesssim 10^{10}~\Msun$ gives a $p$-value of 0.03, suggesting that the differences in $f_{\mathrm{DeV}}$ are statistically significant. In other words, group PSBs are more disk-like than field PSBs, implying that environments might (at least partly) account for the rapid quenching of PSBs in galaxy groups.

To summarize, environmental effects for local PSBs are observable only at $M_*\lesssim 10^{10}~\Msun$; group PSBs are observed to be slightly larger in galaxy size and more disk-like compared to field PSBs. These results are qualitatively consistent with the environment-driven fast quenching scenario described in \cite{pawlik2019diverse}.

\subsection{Influence of the Group Criterion}
\label{subsec:group_criterion}

Our definition of ``group'' is based on the choice of number of members, which is set to $N_{\mathrm{member}}\geq 10$ in this work. However, we also test smaller $N_{\mathrm{member}}$ and find that most of the results about the environments still stand up.

Specifically, we use a more relaxed galaxy group criterion $N_{\mathrm{member}}\geq 5$ to define group PSBs and find that: (1) the overall trend and relative differences between $f_{\mathrm{group}}$ exhibited in Figure \ref{fig:f_group} are nearly unchanged although the absolute values are systematically higher, (2) the only difference of the size--mass relation for group/field PSBs is that more PSBs at $M_*\gtrsim 10^{10.5}~\Msun$ are classified as group PSBs and form a size--mass relation well consistent with that of field PSBs at $10^{10.5}~\Msun \lesssim M_*\lesssim 10^{11}~\Msun$, and (3) no substantial change is found for the $f_{\mathrm{DeV}}$--mass relation.

\subsection{Influence of Dust Effect on Sample Selection Criteria}
\label{subsec:influ_dust}

Giving the aim of selecting galaxies with no ongoing star formation, as stated in Section \ref{subsec:selection}, we adopt a criterion of $\mathrm{EW(\ha)}<3$ for \ha\ emission lines. However, \ha\ emission lines might suffer heavy dust attenuation, leading to the underestimation of the intrinsic fluxes and thus a risk that some SFGs are wrongly classified as PSBs. Given the low S/Ns of the observed emission lines for our PSBs, it is impossible to correct dust attenuation of EW(\ha) via the widely used Balmer decrement method (i..e, the observe $\ha/\hb$ flux ratio). Nevertheless, one can roughly estimate the unattenuated EW(\ha) if the dust attenuation of the underlying stellar continuum is available. More specifically, the ratio between the intrinsic and observed $\mathrm{EW(\ha)}$ can be written as
\begin{equation} \label{eq:logew}
\begin{split}
    \log\left(\frac{\mathrm{EW(\ha)_{int}}}{\mathrm{EW(\ha)_{obs}}}\right)
    &=0.4(A_{\rm H\alpha,gas}-A_{\rm H\alpha,cont})\\
    &=\frac{0.4 k_{\rm H\alpha}}{k_{V}}(A_{V,\rm gas}-A_{V,\rm cont})\\
    &=\frac{0.4 k_{\rm H\alpha} A_{V,\rm cont}}{k_{V}}\left(\frac{1}{R_{\rm att}}-1\right)
\end{split}
\end{equation}
in which $A_{\rm H\alpha,gas}$ ($A_{V,\rm gas}$) and $A_{\rm H\alpha,cont}$ ($A_{V,\rm cont}$) are the dust attenuation at the wavelength of \ha\ ($V$ band) for ionized gas and stellar continua, respectively; $k_{\rm H\alpha}$ and $k_{V}$ are defined via $k_{\lambda}=A_{\lambda}/E(B-V)$), namely the values of the assumed dust extinction/attenuation curve at the wavelengths of \ha\ and $V$ band (i.e., 5500 \AA), respectively; $R_{\rm att}=A_{V,\rm cont}/A_{V,\rm gas}$ is the dust attenuation ratio between ionized gas and stellar continua. Obviously, when dust attenuation of ionized gas is the same as that of stellar continua (i.e., $R_{\rm att}=1$), $\mathrm{EW(\ha)_{int}}=\mathrm{EW(\ha)_{obs}}$, i.e., correction for dust is unnecessary. In this case, our EW(\ha) criterion is unaffected by dust. However, heavier dust attenuation for ionized gas compared to stellar light (i.e., $R_{\rm att}<1$) is reported by observations of SFGs (e.g., \citealt{Calzetti1997AIPC}), while $R_{\rm att}$ is also found to strongly depend on other physical properties, such as $M_*$ \citep{Zahid2017ApJ,Koyama2019PASJ,Lin2020ApJ}.

To roughly evaluate the influence of this effect, we use Equation (\ref{eq:logew}) to estimate the dust correction for EW(\ha). $A_{V,\rm cont}$ can be derived via $\tau_{V,\rm cont}=A_{V,\rm cont}/1.086$ where $\tau_{V,\rm cont}$ is the $V$ band optical depth extracted from the MPA-JHU catalog and derived from the best-fit model of photometry fitting. The $M_*$--$R_{\rm att}$ relation of $R_{\rm att}=3.460-0.277\times\log(M_*/\Msun)$ for local SFGs \citep{Lin2020ApJ} is adopted to obtain $R_{\rm att}$ for each galaxy. An extinction curve of \cite{Cardelli1989ApJ} and $R_{V}=3.1$ are assumed. After $\mathrm{EW(\ha)_{int}}$ is obtained, we use this dust corrected EW(\ha) to redo the sample selection and find that only 388 (26.2\%) galaxies are removed from the original Sample PSB. After carefully remaking figures and tables, we finally find that our main results are nearly unchanged given this dust corrected PSB sample.

Giving the difficulty of correction for dust and the possible large uncertainties of attenuation introduced by this correction, such dust effect is not accounted for in most of previous spectroscopically selected PSB studies (e.g., \citealt{Zahid2016ApJ,chen2019poststarburst}). In order to be in consonance with previous works, we therefore do not correct this effect in our analyses and just represent a short discussion on its influence here.

\subsection{Other Issues Related to Galaxy Size Measurements}
\label{subsec:influ_re}

In this section, we will discuss three issues related to the galaxy size measurements, i.e., the PSF effects, the sampled rest-frame wavelengths, and the systematic uncertainties in $R_{\rm e}$ estimation.

As described in Section \ref{subsec:determination_para} galaxy size measurements used in this work are $R_{\rm e}$ derived from the single S\'ersic fits extracted from the \cite{meert2015catalogue} catalog, which are carried out by \texttt{GALFIT} \citep{peng2002detailed}. Our results reveal that PSB galaxies tend to have extremely small sizes, specially at the low-$M_*$ end. However, the existence of a large PSF will prevent us from obtaining reliable size measurements for very 
compact sources. In \texttt{GALFIT}, the effect of PSF is accounted for during the model fitting. The code uses convolution technique to account for this effect, namely convolves the model image (e.g., a S\'erisic profile) with the PSF before comparing the result with the observed image \citep{peng2002detailed}. \cite{meert2015catalogue} presented a discussion on how the PSF affects size measurement of bulge (i.e., the central compact component of galaxies) and claimed that bulge sizes smaller than 80\% of the half-width at half-maximum (HWHM) of the PSF can be overestimated \citep{Gadotti2008MNRAS}. Such bias might also exist in the single S\'ersic fits for compact galaxies, e.g., our PSB sample.

Given that the HWHM of the PSF for SDSS is about 0.7\arcsec\ in the $r$-band\footnote{\url{https://www.sdss.org/dr12/scope/}}, our further check finds that only 5.2\% of our PSBs have $R_{\rm e}$ smaller than $0.8 \times \mathrm{HWHM}$. This small portion of the PSB sample is concentrated between $10^{10}~\Msun < M_* < 10^{11}~\Msun$ and could not bias the trend at the high- or low-$M_*$ end where the number of PSBs is small. Therefore, the potential PSF effect is too small to have significant influence on our results.

Enabling a direct comparison of size--mass relation with previous PSB studies (e.g., \citealt{yano2016relation, almaini2017massive,wu2018fast,wu2020colors}) requires consistent size measurements with these works in terms of not only the derivation method but also the sampled rest-frame wavelengths since galaxy size might vary with wavelengths. In this work, the former condition is naturally met by design (see Section \ref{subsec:determination_para}). The latter one is also generally satisfied, although misplacement of wavelengths still exists but is proved to have no influence on our main results as discussed below.

The S\'ersic $R_{\rm e}$ we adopted in this work is based on the SDSS $r$-band image, for which the central wavelength corresponds to a rest-frame wavelength of $\sim 5500$ \AA\ given a median redshift of $z=0.13$ and generally covers the rest-frame $g$- and $r$- bands (i.e., $\sim 4600-6200$ \AA) given the redshift range of our PSB sample. Due to the different redshift ranges and images used, the rest-frame wavelengths sampled by previous studies span a large range from $\sim 4100$ \AA\ ($g$-band; e.g., \citealt{wu2018fast,wu2020colors}) to $\sim 8400$ \AA\ ($i$-band; e.g., \citealt{belli2015stellar}). Although all of these works share part of the sampled rest-frame wavelength coverage with our PSB sample, we still note that some galaxies in these samples might move out of the coverage of our sample. We thus further check the main results by using the S\'ersic $R_{\rm e}$ derived based on the SDSS $g$- and $i$-band images, which are also provided by the \cite{meert2015catalogue} catalog, and find that these results are unchanged when $g$- or $i$-band S\'ersic $R_{\rm e}$ are considered. In other words, the observed differences in galaxy sizes between PSBs and QGs can be found from $g$- band to $i$-band, thus our comparison with previous studies are still reasonable.

Finally, we discuss the systematic uncertainties arising from adopting a single S\'ersic profile to describe the light profiles of PSB galaxies. As demonstrated in \cite{Meert2013MNRAS}, the single S\'ersic fit of \texttt{PyMorph} is able to give an unbiased estimation of $R_{\rm e}$ with a small scatter ($\sim 5\%$) for a galaxy with a S\'ersic profile, however, applying the same fit to a galaxy with a two-component S\'ersic + Exponential profile might result in an overestimation of $\sim 5\%$ in $R_{\rm e}$ at the faint end.

We compute the fraction of galaxies that better described by a singe S\'ersic model rather than a S\'ersic + Exponential model for PSBs and all control samples. This fraction of PSBs is found to be comparable with that of Sample Control QG at $M_*\gtrsim 10^{10.5}~\Msun$, but tends to become smaller up to 13\% at the low-$M_*$ end. When we perform a galaxy size comparison between PSBs and QGs, the systematic overestimation of $R_{\rm e}$ of PSBs might be more serious compared to that of QGs, possibly resulting in an underestimation of the differences in $R_{\rm e}$ at the low-$M_*$ end. Therefore, our main results should be unchanged even when the systematic uncertainties in $R_{\rm e}$ estimation is considered.

\section{Summary}
\label{sec:summary}

We present a study of the size--mass relation for local PSB galaxies sample selected from the SDSS DR8. We compile a sample of 1,479 PSB galaxies at 0.012$<z<$0.324. Control samples are constructed to compare PSBs with SFGs and QGs and discuss the possible evolutionary pathways of our PSB galaxies. Our main conclusions are as follows: 

\begin{enumerate}
    \item Due to the limited magnitude of the SDSS spectra, our spectrally selected PSB sample is biased towards the massive end of local PSBs, although less massive PSBs at $M_*\lesssim 10^{10}~\Msun$ are expected to dominate this population based on studies of photometrically selected PSBs.

    \item Whether a $z$--$M_*$ control sample is constructed or not, the sizes of PSBs are smaller than those of QGs in the vast majority of $M_*$ bins we explore, such differences become larger towards the low-$M_*$ end. The $z$--$M_*$--SFR control sample further reinforces the result of an extreme compact morphology for local PSBs. Examination using an independent measurement of concentration or the S\'ersic index also supports these results.
    \item By comparing with predictions of possible PSB evolutionary pathways from cosmological simulations, we demonstrate that the SFG-to-QG fast quenching following a short-lived starburst event (might be induced by major merger) should be the dominated pathway of our PSB sample. 
    \item Local PSBs at $M_*\lesssim 10^{10}~\Msun$ are more clustered than more massive PSBs. Group PSBs are observed to be slightly larger in galaxy size and more disk-like compared to field PSBs, hinting an environment-driven fast quenching pathway for group PSBs, especially at $M_*\lesssim 10^{10}~\Msun$.
\end{enumerate}

The evolutionary pathway of massive PSB galaxies is clearer compared to PSBs at $M_*\lesssim 10^{10}~\Msun$. Although our observations qualitatively support the environment-driven fast quenching scenario described in \cite{pawlik2019diverse}, stronger (quantitative) conclusion about the environmental effect cannot be drawn due to the small number of low-$M_*$ PSBs in our sample. Observations of a large sample of low-$M_*$ PSBs are needed to clarify the role of environments in the evolution of PSB galaxies.

\section*{Acknowledgements}
We would like to thank the referee for many helpful comments and suggestions, which greatly improve the paper. This work is supported by the Strategic Priority Research Program of Chinese Academy of Sciences (No. XDB 41000000), the National Key R\&D Program of China (2017YFA0402600, 2017YFA0402702), the NSFC grant (Nos. 11973038 and 11973039), and the Chinese Space Station Telescope (CSST) Project. HXZ also thanks a support from the CAS Pioneer Hundred Talents Program. Z.S.L acknowledges the support from China Postdoctoral Science Foundation (2021M700137).

This research has made use of the SIMBAD database, operated at CDS, Strasbourg, France. All plots in this paper were built with the {\sc matplotlib} package for {\sc python} \citep{hunter2007matplotlib}.

Funding for SDSS-III has been provided by the Alfred P. Sloan Foundation, the Participating Institutions, the National Science Foundation, and the U.S. Department of Energy Office of Science. The SDSS-III web site is http://www.sdss3.org/.

SDSS-III is managed by the Astrophysical Research Consortium for the Participating Institutions of the SDSS-III Collaboration including the University of Arizona, the Brazilian Participation Group, Brookhaven National Laboratory, Carnegie Mellon University, University of Florida, the French Participation Group, the German Participation Group, Harvard University, the Instituto de Astrofisica de Canarias, the Michigan State/Notre Dame/JINA Participation Group, Johns Hopkins University, Lawrence Berkeley National Laboratory, Max Planck Institute for Astrophysics, Max Planck Institute for Extraterrestrial Physics, New Mexico State University, New York University, Ohio State University, Pennsylvania State University, University of Portsmouth, Princeton University, the Spanish Participation Group, University of Tokyo, University of Utah, Vanderbilt University, University of Virginia, University of Washington, and Yale University.

\bibliography{psb}{}
\bibliographystyle{aasjournal}



\end{document}